\documentclass[twocolumn ]{aastex631}
\shorttitle{MIR M83 nucleus}
\shortauthors{Hernandez et al.}
\graphicspath{{./}{figures/}}

\begin{document}

\title{Dissecting the Mid-Infrared Heart of M83 with JWST}

\author[0000-0003-4857-8699]{Svea Hernandez}
\affiliation{AURA for ESA, Space Telescope Science Institute, 3700 San Martin Drive, Baltimore, MD 21218, USA}

\author{Logan Jones}
\affiliation{Space Telescope Science Institute,
3700 San Martin Drive, 
Baltimore, MD 21218, USA}

\author[0000-0002-0806-168X]{Linda J. Smith}
\affiliation{Space Telescope Science Institute,
3700 San Martin Drive, 
Baltimore, MD 21218, USA}

\author[0000-0001-5042-3421]{Aditya Togi}
\affiliation{Texas State University, Department of Physics, 601 University Dr, San Marcos, TX 78666, USA}

\author[0000-0003-4137-882X]{Alessandra Aloisi}
\affiliation{Space Telescope Science Institute,
3700 San Martin Drive, 
Baltimore, MD 21218, USA}

\author[0000-0003-2379-6518]{William P. Blair}
\affil{The William H. Miller III Department of Physics and Astronomy, 
Johns Hopkins University, 3400 N. Charles Street, Baltimore, MD, 21218; 
wblair@jhu.edu}

\author[0000-0002-2954-8622]{Alec S.\ Hirschauer}
\affil{Space Telescope Science Institute, 3700 San Martin Drive, Baltimore, MD 21218, USA}

\author{Leslie K. Hunt}
\affil{INAF - Osservatorio Astrofisico di Arcetri, Largo E. Fermi 5, 50125 Firenze, Italy}

\author[0000-0003-4372-2006]{Bethan L.\ James}
\affil{AURA for ESA, Space Telescope Science Institute, 3700 San Martin Drive, Baltimore, MD 21218, USA}

\author{Nimisha Kumari}
\affil{AURA for ESA, Space Telescope Science Institute, 3700 San Martin Drive, Baltimore, MD 21218, USA}

\author[0000-0002-7716-6223]{Vianney Lebouteiller}
\affil{AIM, CEA, CNRS, Université Paris-Saclay, Université
Paris Diderot, Sorbonne Paris Cité, F-91191 Gif-sur-Yvette, France}

\author[0000-0003-2589-762X]{Matilde Mingozzi}
\affiliation{Space Telescope Science Institute, 3700 San Martin Drive, Baltimore, MD 21218, USA}

\author[0000-0002-9190-9986]{Lise Ramambason}
\affil{Institut fur Theoretische Astrophysik, Zentrum für Astronomie, Universität Heidelberg, Albert-Ueberle-Str. 2, D-69120 Heidelberg, Germany}



\begin{abstract}
We present a first look at the MRS observations of the nucleus of the nearby galaxy M83, taken with MIRI onboard JWST. The observations show a rich set of emission features from the ionized gas, warm molecular gas, and dust. To begin dissecting the complex processes in this part of the galaxy, we divide the observations into four different regions. We find that the strength of the emission features varies strongly from region to region, with the south-east region displaying the weakest features tracing the dust continuum and ISM properties. Comparison between the cold molecular gas traced by the $^{12}$CO (1-0) transition with ALMA and the H$_2$ S(1) transition shows a similar spatial distribution. This is in contrast to the distribution of the much warmer H$_2$ emission from the S(7) transition found to be concentrated around the optical nucleus. We use the rotational emission lines and model the H$_2$ excitation to estimate a total molecular gas mass accounting for the warm H$_2$ component of M($>$50 K)$_{\rm H_{2}}$ = 67.90 ($\pm 5.43$)$\times$10$^{6}$ M$_{\odot}$. We compare this value to the total gas mass inferred by probing the cold H$_2$ gas through the $^{12}$CO (1-0) emission, M(CO)$_{\rm H_{2}}$ = 17.15$\times$10$^{6}$ M$_{\odot}$. We estimate that $\sim$75\% of the total molecular gas mass is contained in the warm H$_2$ component. We also identify [\ion{O}{4}] 25.89 $\mu$m and [\ion{Fe}{2}] 25.99 $\mu$m emission. We propose that the diffuse [\ion{Fe}{2}] 25.99 $\mu$m emission might be tracing shocks created during the interactions between the hot wind produced by the starburst and the much cooler ISM above the galactic plane. More detailed studies are needed to confirm such a scenario.

\end{abstract}



\section{Introduction} \label{sec:intro}
Starburst galaxies are ideal laboratories for studying the physics of star formation (SF) and overall galaxy evolution. These galaxies are commonly forming stars at a much higher rate than typical star-forming systems \citep[e.g., SF rate per unit area $I_{\rm SF}$ = 1-100 M$_\odot$ yr$^{-1}$ kpc$^{-2}$;][]{hec94, meu97}. Overall, starburst galaxies cover a broad range in morphology from blue compact dwarfs (BCDs, which share common properties with the high-$z$ environments; e.g., \citealt{izo21}) to spiral galaxies and merging systems \citep{lei95, lei95b, ann20}. Starbursts are believed to be a critical component in the evolution of galaxies as these are connected to processes such as the fueling of active galactic nuclei \citep[AGN;][]{kna04}, production of large populations of massive stars and massive star clusters \citep{meu95, ho96, ada10, zha18}, and powerful stellar/AGN outflows \citep{hec90, rup18, cai22, yux22}. Due to the nature of starburst galaxies, several studies have suggested \citep{rob13, rob15, fin19}, and confirmed, that these compact systems could have been dominant contributors in the reionization of the Universe \citep{mad91, hec11, izo18b, izo18, flu22, flu22b, mar22}. Furthermore, these systems are believed to be responsible for $\sim$25\% of the star-formation rate on the high-mass end in the local universe, and dominate the star-forming activity at high redshift \citep[$z > 0.7$; ][]{hec98, flo05}. Starbursts in these star-forming systems are generally located in smaller regions of a galaxy; in the case of large spirals these are located in the very central regions and are known as starburst nuclei \citep[e.g.,][]{bal83}. \par
Previous generations of infrared (IR) telescopes such as the Infrared Space Observatory (ISO) and the more recent Spitzer Space Telescope \citep{wer04} have highlighted the complexity and richness of the mid-IR (MIR) observations of starburst nuclei \citep{gen00, arm04, arm06, lah07, arm20}. Studies analyzing spectroscopic observations taken with the Infrared Spectrograph \citep[IRS; ][]{hou04} on Spitzer uncovered strong atomic emission lines, many arising from highly-ionized gas (e.g., [\ion{Ne}{5}]; \citealt{arm07}), polycyclic aromatic hydrocarbons (PAH) emission from dust grains \citep{bra06, smi07}, strong emission from the warm molecular gas \citep[H$_2$;][]{arm06, app06, sti14}, and intense [\ion{Fe}{2}] emission indicating the presence of shocks \citep{ina13}. The MIR understanding of starburst nuclei, and starbursting galaxies in general, will be revolutionized with the recent arrival of the James Webb Space Telescope (JWST). The medium resolution spectrometer (MRS) on the JWST Mid-Infrared Instrument (MIRI) is already allowing for studies of the detailed properties of these objects at spatial scales seven times smaller, fifty times more sensitive, and with spectral resolutions five times higher than those provided by the IRS on Spitzer \citep{rie15, lib21, san22, ada22,thi23}. \par

At a distance of 4.6 Mpc \citep{sah06}, M83 is the nearest face-on spiral galaxy. The confirmed central starburst \citep{dia06, dop10}, its orientation, and its metallicity ($\sim$ 1 -- 3 Z$_{\odot}$, with a high level of chemical enrichment e.g., \citealt{bre02, bre16, her21}), make this galaxy an ideal environment to study in detail the MIR emission properties of starburst nuclei. The interest in the nucleus of M83 has increased as past studies have suggested that the ongoing starburst was most likely triggered by a recent minor merger \citep{kna10}. This spiral galaxy has been studied extensively at a broad range of wavelengths: ultraviolet \citep[UV; ][]{jam14, her19, her21, bru20}, optical \citep{mas06, kim12, bla14}, IR \citep{tha00, kna10, wil15, wu15}, radio \citep[][]{mad06, rus20}, and X-ray \citep{duc13, lon14,hun21,wan21}. The starburst nucleus has shown to be a complex system with a reported double ring structure \citep[e.g., ][]{elm98}. Additionally, through analysis of three-dimensional spectroscopy data in the R band, studies have found that the nucleus of the galaxy is offset from its photometric and kinematic center by 4\arcsec, which has been interpreted as evidence for a second nucleus \citep{tha00, sak04,mas06, dia06, rod09} or as the result of a perturbation in the gravitational potential from a possible past interaction \citep{hou08}. \par
Being a site of vigorous star formation, the nucleus of M83 is an excellent object to be studied in the MIR to uncover the processes fueling star formation and the complex interplay between star formation, ionized gas, and dust (both in emission and absorption). Here we report on new MIR integral field spectroscopy of the core of M83 taken with the MRS mode of MIRI onboard JWST and focus on two specific components: (1) the molecular gas component, and (2) possible evidence and origin of shocks. We highlight that future studies will further explore the MIRI/MRS observations expanding on the physical and chemical properties of the multi-phase ISM and dust components in the nuclear regions of M83 (Jones et al. in prep.). This paper is structured as follows: In Section \ref{sec:obs} we present a description of the new MIRI/MRS observations and data reduction, in Section \ref{sec:results} we describe the mid-infrared properties of the core of M83, and in Sections \ref{sec:discussion} and \ref{sec:summary} we present a discussion of the observed properties and a brief summary, respectively. As an initial note, we highlight that all of the molecular gas masses discussed in this work refer to the H$_2$ masses, with no heavy element correction. 

\section{Observations and Data Reduction} \label{sec:obs}
\begin{figure*}
   	  \centerline{\includegraphics[scale=0.25]{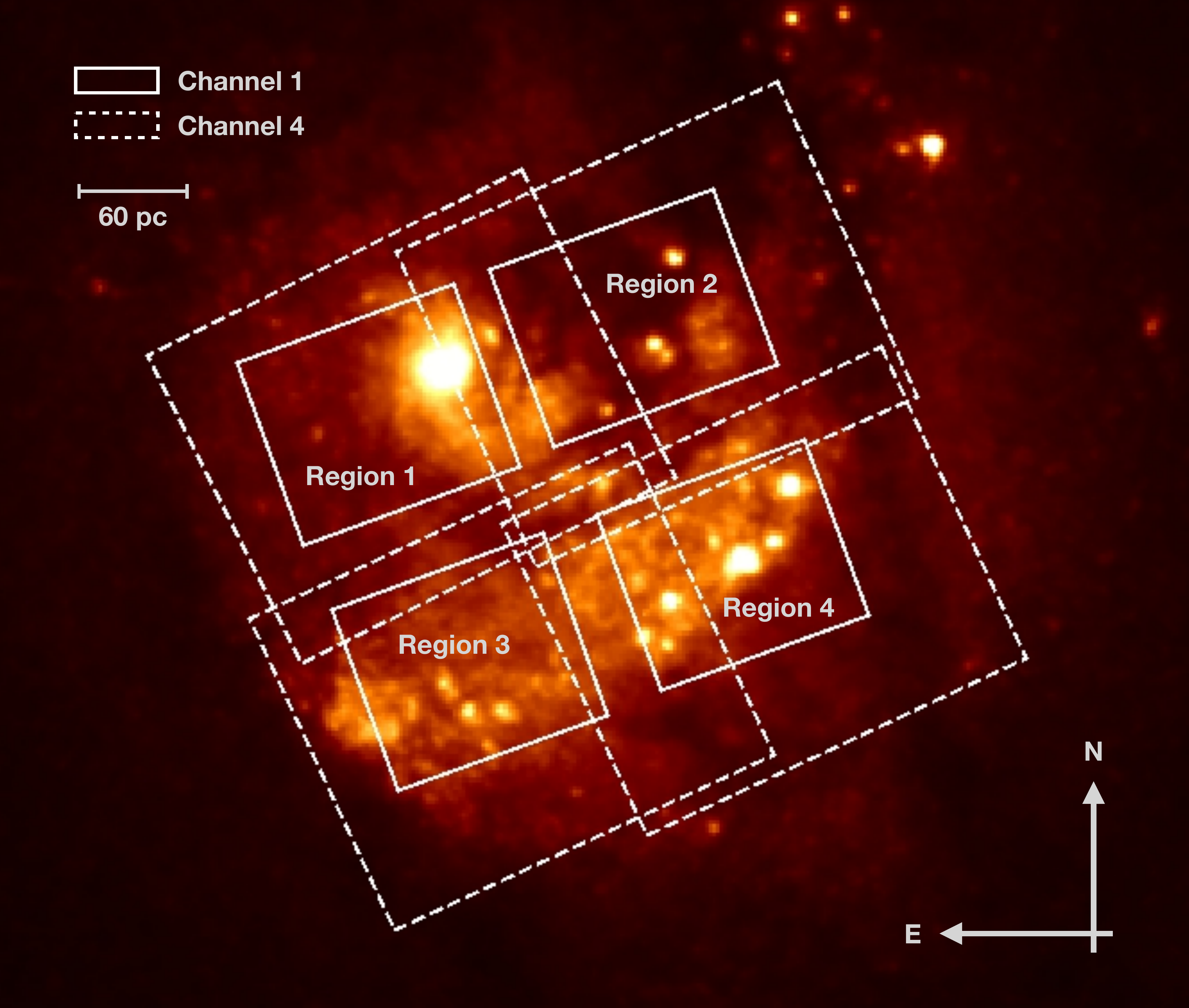}}
      \caption{HST/WFC3 (F110W, resolution $\sim$0.11\arcsec, PID: 11360) image of the core of M83. We show with white boxes the location of the JWST MIRI/MRS Channel 1 (solid) with coverage at short wavelengths (5-8 $\micron$) and Channel 4 (dashed), with full coverage at longer wavelengths (17-28 $\micron$). The labels in each quadrant are the adopted regions throughout our analysis.}
         \label{fig:m83_miri_wfc3}
   \end{figure*}

The JWST observations presented in this paper were collected as part of Cycle 1 under PID 02219 (PI: Hernandez) between July 5 and 9, 2022. Given the expected pointing accuracy of JWST ($\sim$0.10\arcsec), the program did not require a target acquisition. The M83 observations were taken with the MIRI/MRS instrument to create a 2$\times$2 mosaic which fully sampled the entire nucleus of this nearby galaxy primarily with the largest MRS channel 4 (with a field of view, FOV, of 6.9\arcsec$\times$7.9\arcsec per pointing, see Figure \ref{fig:m83_miri_wfc3}). A 4-point dither pattern optimized for extended sources was also applied to achieve optimal sampling throughout the MRS FOV and to identify and remove detector artifacts. The individual pointings were strategically positioned to fully sample the massive stellar clusters with all four MIRI/MRS channels. To avoid data excess problems the observations were divided into three separate visits, each observing with a different grating, SHORT (A), MEDIUM (B), LONG (C), to obtain contiguous wavelength coverage from $\sim$5 to 28 $\micron$. Each pointing was observed with 40 groups per integration for a total of five integrations and a total exposure time of 2264 sec. Additionally, to accurately measure and correct for the thermal background, the program collected companion background MRS cubes avoiding contamination from the disk of the galaxy with identical integrations as those of the individual pointings. \par
The MIRI/MRS raw products were retrieved from the Mikulski Archive for Space Telescopes (MAST). The observations were reduced with the JWST calibration software version (\texttt{CAL\_VER}) 1.6.2. We also applied residual fringe corrections available in the JWST pipeline to both the three-dimensional and one-dimensional observations, in the latter taking care of the low-level dichroic fringing pattern. All the {\it JWST} data used in this paper can be found in MAST: \dataset[10.17909/a61h-f081]{http://dx.doi.org/10.17909/a61h-f081}.

\section{Results} \label{sec:results}
\begin{figure*}
   	  \centerline{\includegraphics[scale=0.33]{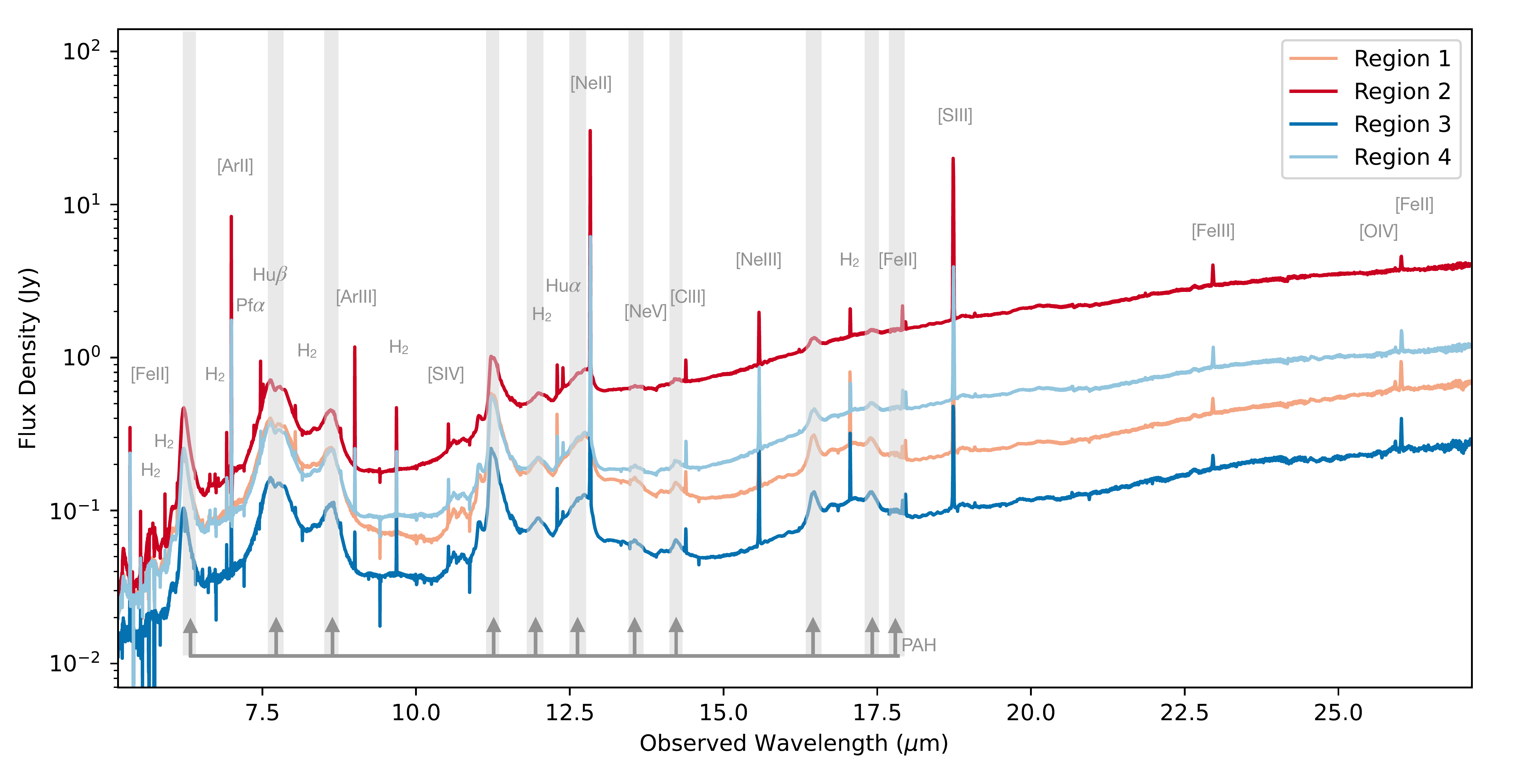}}
      \caption{Mid-infrared spectra of four different regions in the nucleus of M83 as observed by JWST/MIRI. Note that the spectra from the MIRI/MRS Channels 2-3 have been scaled to match those from Channel 1. We label the strongest emission lines present, including PAH features.}
         \label{fig:m83_regions_spec}
   \end{figure*}

Our MIRI/MRS observations roughly cover the core of M83 in a region of approximately 200 pc $\times$ 200 pc. To begin dissecting the complex processes taking place in this part of the galaxy, we divided the nucleus of M83 into four different regions primarily following the individual MIRI/MRS pointings (see Figure \ref{fig:m83_miri_wfc3}). We extracted a single spectrum per region (per channel) and correct for foreground extinction adopting the latest MIR extinction curve by \citet{gor21}. Given that the different channels have slightly different FOVs, we scaled the extracted spectra for channels 2, 3 and 4 to the continuum of those from channel 1 between 5.57 and 5.59 $\micron$, a particularly featureless wavelength range. The corrected spectra analyzed in this work are shown in Figure \ref{fig:m83_regions_spec}. \par
All four regions in the nucleus of M83 exhibit both strong and weak PAH emission features (6.2, 7.7, 8.6, 11.3, 12.0, 12.7, 13.5, 14.2, 16.4, 17.4, 17.8 $\micron$, highlighted by the bottom grey arrows in Figure \ref{fig:m83_regions_spec}). We list in Table \ref{table:pah} the inferred fluxes for some of the strongest PAH features in the different regions. These initial measurements were done with the \texttt{PAHFIT} software \citep{smi07}. The spectra are also rich in emission from fine-structure lines such as [\ion{Fe}{2}] 5.34\micron, [\ion{Fe}{3}] 22.93\micron, [\ion{Fe}{2}] 25.99\micron, [\ion{Ar}{2}] 6.99\micron, [\ion{Ar}{3}] 8.99\micron, [\ion{S}{3}] 18.71\micron, [\ion{S}{4}] 10.51\micron, [\ion{Ne}{2}] 12.81\micron, [\ion{Ne}{3}] 15.56\micron, [\ion{Cl}{2}] 14.37\micron, and [\ion{O}{4}] 25.89\micron, as well as hydrogen recombination lines (Pf$\alpha$ 7.46\micron, Hu$\beta$ 7.50\micron, Hu$\alpha$ 12.37\micron) with strengths varying from region to region. In Table \ref{table:fluxes} we list the foreground extinction-corrected fluxes for a broad range of emission lines, along with their corresponding full width at half maximum (FWHM). Finally, we also detect emission from seven H$_{2}$ 0-0 rotational lines, S(1) through S(7), and similar to the fine-structure lines, the emission from the warm molecular gas appears to vary strongly from region to region (see Table \ref{table:h2} for a list of foreground extinction-corrected  H$_2$ flux measurements). Specifically for these H$_2$ lines we observe velocity shifts of the order of $\sim$100 km s$^{-1}$ from Region 1 to region 4, comparable to values in the the velocity maps by \citet{cal21}. Many of these lines have been previously detected in Spitzer/IRS spectra (e.g., PI: Rieke, PID: 59); however, the higher spectral resolution of the MIRI/MRS observations allow us to analyze the distribution and dynamics of the molecular and atomic gas, as well as dust, in a spatially-resolved manner at spatial scales of $\sim$100 pc in the present study, and we plan to explore the physical properties of these regions at the spatial scales of a few tens of pc in future studies.

       \begin{figure*}
   	  \centerline{\includegraphics[scale=0.56]{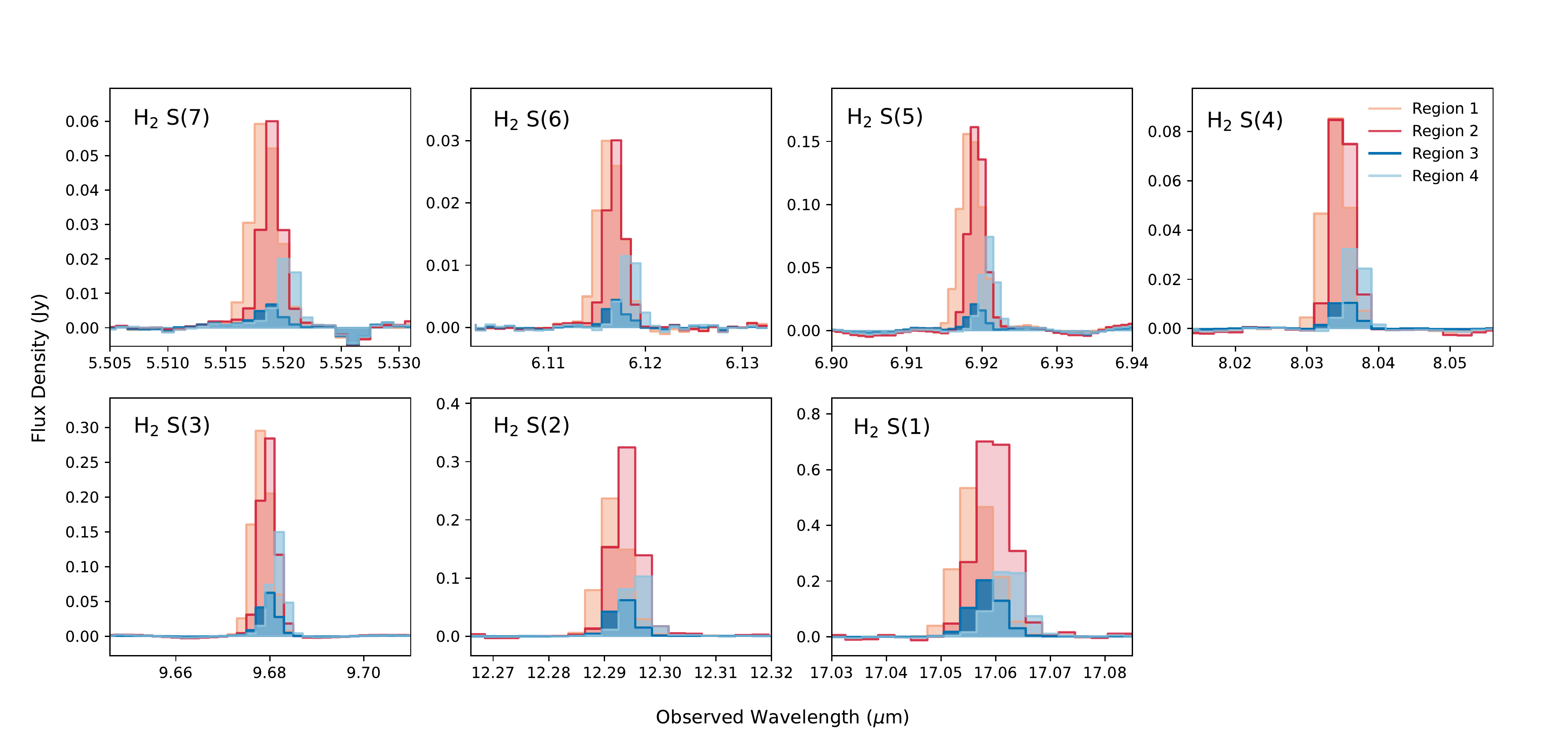}}
      \caption{\textbf{Foreground} extinction-corrected continuum-subtracted H$_{2}$ emission for seven different transitions detected in four regions in the core of M83. The H$_{2}$ profiles are shown at the observed wavelengths.}
         \label{fig:m83_H2}
   \end{figure*}
   
                \begin{figure}
   	  \centerline{\includegraphics[scale=0.3]{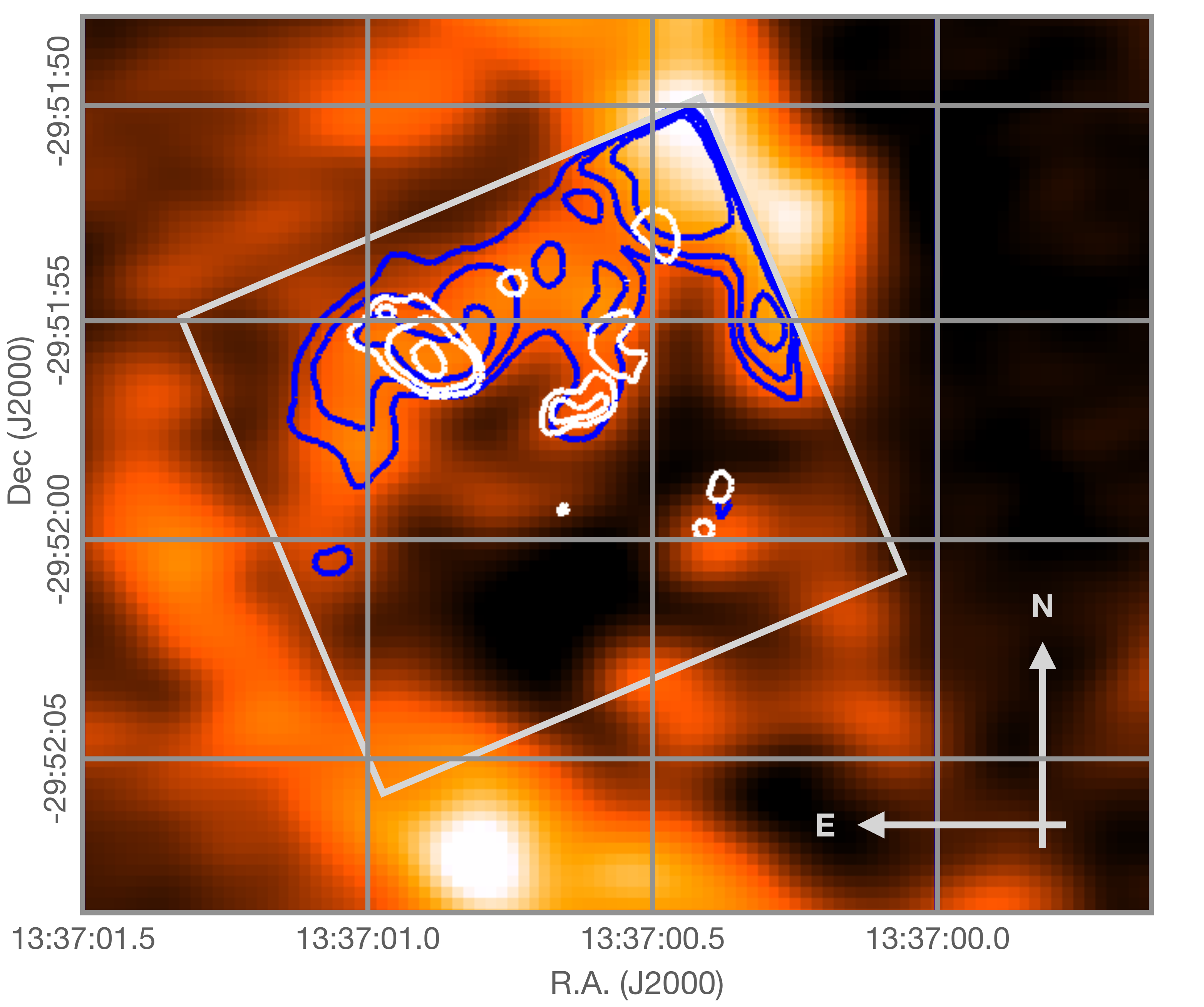}}
      \caption{$^{12}$CO (1--0) emission map of the nucleus of M83 by \citet{hir18}. The combined CO map has a resolution of 2.03\arcsec $\times$ 1.15\arcsec. We show with a grey box the FoV of the combined MIRI/MRS channel-3  pointings. The blue contours show the concentrations of H$_2$ S(1) 17.035 $\micron$ emission, compared to the distribution of the much warmer H$_2$ from the S(7) 5.511 $\micron$ transition shown in white contours. }
         \label{fig:CO_H2}
   \end{figure}

          \begin{figure}
   	  \centerline{\includegraphics[scale=0.38]{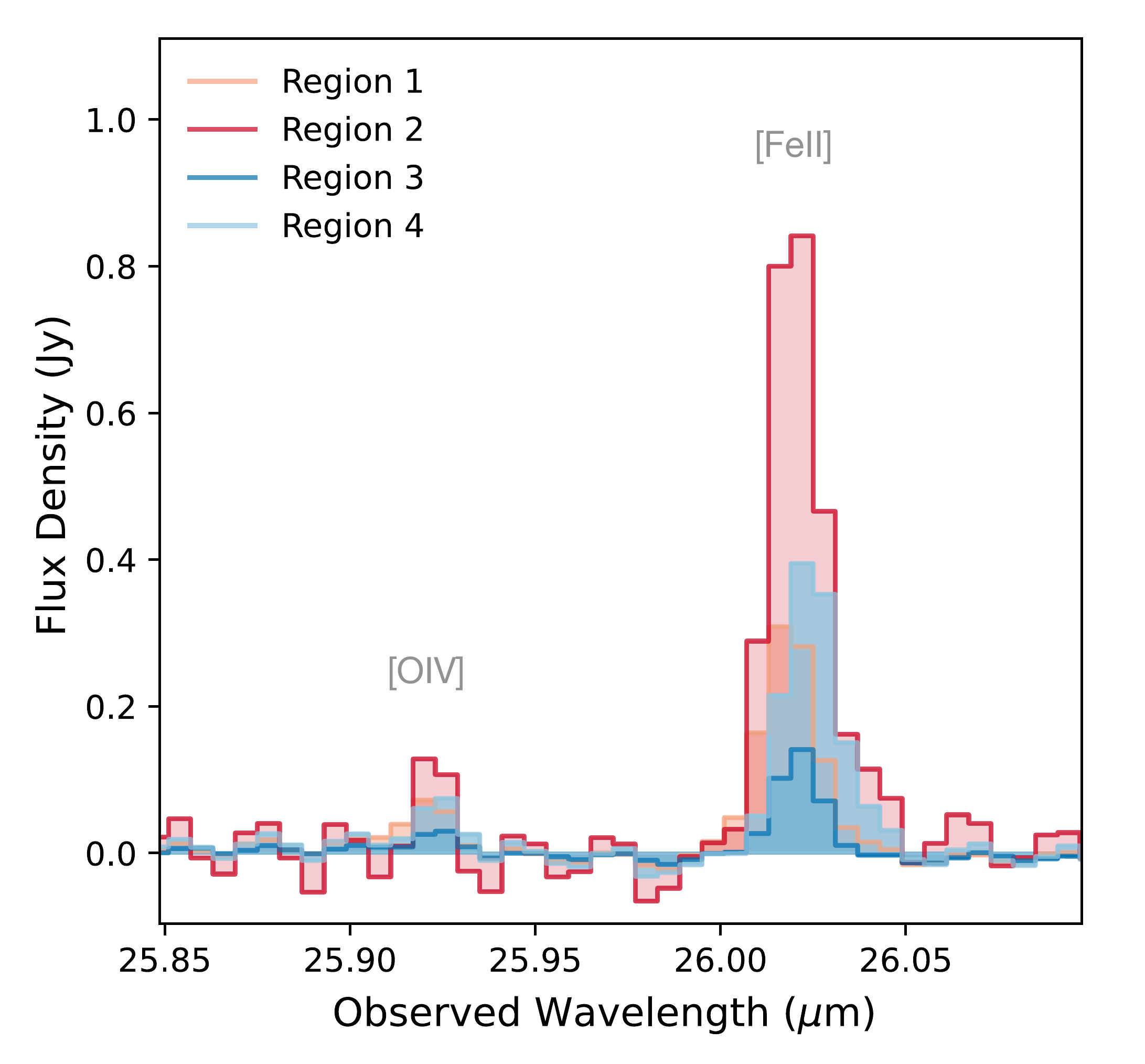}}
      \caption{Continuum-subtracted [OIV] 25.89 $\mu$m and [FeII] 25.99 $\mu$m emission detected in all four regions in the core of M83. }
         \label{fig:m83_oiv_feii}
   \end{figure}
   
   \begin{table*}
\caption{PAH emission line fluxes}
\label{table:pah}
\centering 
\begin{tabular}{crrrrr}
\hline \hline
\multicolumn{2}{c}{Line} &  \multicolumn{1}{c}{Region 1} &  \multicolumn{1}{c}{Region 2} &  \multicolumn{1}{c}{Region 3} &  \multicolumn{1}{c}{Region 4} \\
\multicolumn{2}{c}{Rest Wavelength} & \multicolumn{4}{c}{Flux} \\
\multicolumn{2}{c}{($\mu$m)} & \multicolumn{4}{c}{(10$^{-16}$ $W\; m^{-2}$)} \\
\hline
PAH & 6.22 & 43.30$\pm$0.15 & 79.20$\pm$0.17  & 16.40$\pm$0.11 & 38.80$\pm$0.13\\
PAH  &7.42 & 34.40$\pm$0.58 & 59.10$\pm$0.52& 4.03$\pm$0.32 & 12.20$\pm$0.41\\
PAH  &7.60 & 59.90$\pm$0.28 & 112.00$\pm$0.30 & 25.50$\pm$0.17 & 60.00$\pm$0.22\\
PAH  &7.85 & 63.50$\pm$0.20 & 104.00$\pm$0.24 & 24.70$\pm$0.12 & 51.50$\pm$0.17\\
PAH  &8.61 & 34.50$\pm$0.12& 56.60$\pm$0.13 & 14.00$\pm$0.10 & 30.30$\pm$0.91\\
PAH  &11.33 & 29.50$\pm$0.15 & 42.70$\pm$0.15 & 10.50$\pm$0.65 & 23.30$\pm$0.92\\
PAH  &12.62 & 25.50$\pm$0.09 & 49.20$\pm$0.15 & 9.69$\pm$0.61 & 21.20$\pm$0.95\\
\hline
\end{tabular}
\end{table*}

	\movetabledown=3.5in	
	\begin{rotatetable*}
\begin{deluxetable}{crrrrrrrrr}
\tablecaption{Mid-infrared emission line properties}\label{table:fluxes}
\tablehead{ \multicolumn{2}{c}{Line} &  \multicolumn{2}{c}{Region 1} &  \multicolumn{2}{c}{Region 2} &  \multicolumn{2}{c}{Region 3} &  \multicolumn{2}{c}{Region 4} \\
\multicolumn{2}{c}{Rest Wavelength} & \multicolumn{1}{c}{Flux} & \multicolumn{1}{c}{FWHM} & \multicolumn{1}{c}{Flux} & \multicolumn{1}{c}{FWHM} & \multicolumn{1}{c}{Flux} & \multicolumn{1}{c}{FWHM} & \multicolumn{1}{c}{Flux} & \multicolumn{1}{c}{FWHM}\\
\multicolumn{2}{c}{($\mu$m)} & \multicolumn{1}{c}{(10$^{-18}$ $W\; m^{-2}$)} & \multicolumn{1}{c}{(km s$^{-1}$)} & \multicolumn{1}{c}{(10$^{-18}$ $W\; m^{-2}$)} & \multicolumn{1}{c}{(km s$^{-1}$)} & \multicolumn{1}{c}{(10$^{-18}$ $W\; m^{-2}$)} & \multicolumn{1}{c}{(km s$^{-1}$)} & \multicolumn{1}{c}{(10$^{-18}$ $W\; m^{-2}$)} & \multicolumn{1}{c}{(km s$^{-1}$)}\\}
\startdata
{[ArII]} &6.99 & 103.81$\pm$2.52 & 168 $\pm$ 2 &  1454.70$\pm$61.92&125 $\pm$ 2 &      44.98$\pm$2.38&145 $\pm$ 4 &      315.85$\pm$1.74 &130 $\pm$ 3 \\
Pf$\alpha$& 7.46  & 5.32$\pm$0.32& 159 $\pm$ 5 &  87.02$\pm$0.01 & 117 $\pm$ 2 &          3.46$\pm$0.01& 167 $\pm$ 10 &    17.99$\pm$1.23& 121 $\pm$ 4\\
Hu$\beta$ &7.50 &2.08$\pm$0.27 &  173 $\pm$ 14 &    23.36$\pm$3.70& 120 $\pm$ 10 &          1.53$\pm$0.25&280 $\pm$ 29 &       6.81$\pm$1.09&163 $\pm$ 14\\
{[ArIII]} & 8.99 & 11.73$\pm$0.36 &   171 $\pm$ 3 &     161.13$\pm$2.50 &145 $\pm$ 1 &          6.74$\pm$0.23&167 $\pm$ 3 &        27.11$\pm$0.57& 148 $\pm$ 2\\
{[SIV]}& 10.51 & 6.11$\pm$0.10 &   165 $\pm$ 4 &    17.21$\pm$1.69 &168 $\pm$ 6 &          2.26$\pm$0.01& 153 $\pm$ 5 &       8.76$\pm$0.74& 164 $\pm$ 8\\
Hu$\alpha$ & 12.37 & 2.32$\pm$0.22 &  152 $\pm$ 8 &      25.76$\pm$1.52 &138 $\pm$ 4 &        0.99$\pm$0.07 &128 $\pm$ 5 &       6.53$\pm$0.51&153 $\pm$ 6 \\
{[NeII]}&  12.81 & 308.49$\pm$5.39 &179 $\pm$ 1 &      3409.99$\pm$127.81 &150 $\pm$ 2 &        131.59$\pm$5.50 &149 $\pm$ 3 &        749.27$\pm$23.80& 162 $\pm$ 2\\
{[NeV]}&  14.32 & 1.15$\pm$0.22 &  342 $\pm$ 36 &       -- & -- & 0.24$\pm$0.10 &197 $\pm$ 38 &       -- & -- \\
{[ClII]} & 14.37  & 6.03$\pm$0.12&  167 $\pm$ 2 &        29.91$\pm$0.97 & 141 $\pm$ 2 &        2.52$\pm$0.08&140 $\pm$ 2 &               10.98$\pm$0.40 & 157 $\pm$ 2\\
{[NeIII]}& 15.56 & 72.13$\pm$1.16 &   199 $\pm$ 2 &       153.19$\pm$2.01 &199 $\pm$ 2 &         30.93$\pm$0.39&175 $\pm$ 2 &             168.91$\pm$2.70&199 $\pm$ 2 \\
{[FeII]}& 17.94& 9.24$\pm$2.09 &  260 $\pm$ 33 &      26.25$\pm$18.50 & 235 $\pm$ 113 &       3.56$\pm$0.71&220 $\pm$ 24 &              14.98$\pm$5.03& 254 $\pm$ 50\\
{[SIII]} & 18.71 & 125.98$\pm$1.04 & 186 $\pm$ 1 &       1768.01$\pm$17.00 & 180 $\pm$ 1 &       38.33$\pm$0.59&184 $\pm$ 1 &             230.53$\pm$6.64& 176 $\pm$ 1 \\
{[FeIII]} & 22.93  & 9.72$\pm$0.69&  202 $\pm$ 8 &      93.90$\pm$4.81 &195 $\pm$ 5 &       3.99$\pm$0.40& 206 $\pm$ 11 &             26.51$\pm$1.41& 192 $\pm$ 5 \\
{[OIV]} & 25.89 & 5.11$\pm$1.18 & 182 $\pm$ 25 &        6.26$\pm$5.37 & 112 $\pm$ 82 &      1.98$\pm$0.60&150 $\pm$ 27 &             4.62$\pm$1.74&146 $\pm$ 33 \\
{[FeII]} & 25.99 & 25.90$\pm$2.20 &  197 $\pm$ 9 &       69.73$\pm$6.25 & 191 $\pm$ 9 &     9.88$\pm$0.96&165 $\pm$ 9 &                  32.58$\pm$2.73& 196 $\pm$ 9\\
\enddata		
\end{deluxetable}
\end{rotatetable*}

\section{Discussion} \label{sec:discussion}
\subsection{Importance of warm H$_2$}\label{sec:h2}
The vast number of stars in the Universe formed primarily from molecular clouds, directly connecting SF to the surface density of molecular gas \citep[e.g.,][]{ken98}. Being such a critical ingredient of SF, it is essential to accurately measure and map the distribution of H$_2$ to fully understand the processes involved in forming stars, and how these evolve over cosmic history. Despite being the most abundant molecule in the universe, given the intrinsic nature of H$_2$ (being a weak rotational emitter), molecular hydrogen can be difficult to study directly. Instead, the next most abundant molecule, CO, is generally used as a molecular gas tracer in the local Universe \citep{ ler11, bol13, sai17}, as well as in intermediate-to-high redshift galaxies \citep[$z\sim$1-2, e.g.,][]{tac10, pav18, tac20}. A conversion factor ($\alpha_{CO}$) is generally needed to measure the total molecular gas masses \citep[e.g.,][]{bol13} as CO emission is consistently found to be optically thick. Recent evidence, indicates that $\alpha_{CO}$ varies substantially within and across galaxies at different epochs and metallicities \citep{san13, isr20}. \par

Additional complications in measuring the total mass of H$_{2}$ involve scenarios where far-ultraviolet (FUV) photons permeate into molecular clouds photodissociating CO molecules, leaving a C$^{+}$-emitting shell around the CO cloud \citep{mad20}. We note that the C$^{+}$-emitting region is larger in those environments where the dust-to-gas ratio decreases and these FUV photons are able to permeate deeper into the cloud. This is in contrast to H$_2$, which photodissociates via absorption of Lyman-Werner band photons, and can become self-shielded from photodissociating \citep{dra11,kru13,gne14}, leaving a potentially significant reservoir of molecular hydrogen outside of the CO-emitting region. CO-dark molecular gas \citep{pol95, mad97, gre05, wol10, ack12, glo12, mad20}, as the name indicates, is not probed by CO, requiring other tracers to accurately quantify the total molecular gas mass, 
\begin{equation}
\rm M(TOTAL)_{\rm H_2} = M(CO)_{\rm H_2} + M(CO-dark)_{\rm H_2}
\end{equation}
where $\rm M(TOTAL)_{\rm H_2}$ is the total molecular gas mass, $\rm M(CO)_{\rm H_2}$ is the molecular gas mass estimated using CO as a tracer of H$_2$, and $\rm M(CO-dark)_{\rm H_2}$ is the molecular gas mass contained in the warmer H$_2$ gas reservoir. \par
Star-forming galaxies (SFGs) typically have massive star clusters capable of photodissociating substantial fractions of CO molecules. Interestingly, theoretical and observational studies have also shown that efficient SF is possible in these CO-dark regions, as well as in regions without H$_2$, suggesting the existence of pathways to forming stars directly in the atomic ISM phase \citep{kru11, glo12b, kum20}. \par

\citet{ada15} have identified a relatively higher cluster formation efficiency in the nucleus of M83, compared to regions outside, which allows for a high enough production of energetic far-ultraviolet (FUV; 6 eV $<$ h$\nu$ $<$ 13.6 eV) photons capable of permeating the molecular clouds and photo-dissociating the CO gas. Given the findings by \citet{ada15}, studies such as that by \citet{her21} have predicted the presence of large quantities of warm molecular gas (CO-dark gas) in the center of this spiral galaxy. The predictions by \citet{her21} come from an observed excess in the column densities measured from the \ion{S}{2} absorption localized to the nuclear regions, and absent on the disk of the galaxy. \citet{her21} proposed that the observed S$^{+}$ excess is due to a significant fraction of the measured \ion{S}{2} tracing the CO-dark (C$^+$-emitting) gas along the line of sight, instead of only tracing the neutral gas. These new MIRI/MRS observations allow us to confirm such a prediction, as well as study the spatial distribution of this warm ($T\gtrsim$ 100 K) molecular gas compared to the much cooler ($\sim$10-20 K) $^{12}$CO (1-0) gas. \par
The temperature equivalent of the upper energy level for the lowest pure rotational line transition, H$_2$ 0-0 S(0) at 28.22$\micron$, is at 510 K. \citet{tog16} note, however, that the Boltzmann distribution of energy levels can lead to excitation even at more conservative temperatures. Such a case might apply to the lowest transitions, S(0) to S(2), where these can be comfortably populated at excitation temperatures of $\sim$150 K, and even as low as $\sim$80 K. 
We detect seven H$_{2}$ transitions in all four regions of the M83 nucleus with varying strengths. In Figure \ref{fig:m83_H2} we show in different panels the H$_{2}$ profiles for each transition color-coded based on the specific region. Although we observe H$_2$ emission from all seven transitions throughout the core of the galaxy, the strongest emission is found in Regions 1 and 2, covering the northern-most areas, including the optical nucleus. Overall, the distribution of warm H$_{2}$ appears to weaken towards the southern regions of the nucleus, with Region 3 exhibiting the weakest H$_{2}$ emission. \par
In Figure \ref{fig:CO_H2} we show the cold molecular gas traced by the $^{12}$CO (1-0) transition with the Atacama Large Millimeter/submillimeter Array (ALMA) as published by \citet{hir18}. Figure \ref{fig:CO_H2} provides a qualitative comparison between the CO emission and the distribution of H$_2$ gas for the S(1) and S(7) transitions in blue and white contours, respectively. The S(1) transition, expected to have the lowest excitation temperature out of all the detected transitions, appears to comfortably follow the cooler CO emission. The emission of the H$_2$ gas with the warmest temperatures, the S(7) transition, is primarily concentrated near the center of M83, covering an area with a diameter of $\sim$60 pc. \par
We adopt the continuous-temperature model by \citet{tog16} along with the measured H$_2$ fluxes to infer the molecular gas mass (no heavy element correction) in the different regions. Assuming the H$_2$ emission is optically thin, the H$_2$ excitation can be modeled through excitation diagrams relating the column density of the upper level of a particular transition ($N_u$) to its energy level ($E_u$), providing warm molecular gas masses by only varying the slope of the power law ($n$) and the lower temperature (T$_l$). Similar to the work by \citet{tog16}, we fixed the upper temperature (T$_u$) of the distribution to 2000 K, however, we highlight that we also tested setting this parameter free and confirmed that the fraction of the total warm molecular gas mass with T$_{u} >$ 2000 K is negligible. In Table \ref{table:h2} we list the inferred foreground extinction-corrected fluxes for the detected transitions used in calculating the H$_2$ masses. \par
We show in columns 2 and 3 of Table \ref{table:h2_model} the two model-derived parameters, $n$ and T$_l$, for all four nuclear regions studied here. We also include in Appendix \ref{sec:app_excitation} the excitation diagrams, along with the best fit models, for each region. The inferred power law indices range between 5.17 (in region 1 containing the optical nucleus) and 6.33. We note that in spite of the differences in spatial scales probed, these indices are within the range observed in the galaxy sample by \citet{tog16}, 3.79 $<$ $n$ $<$ 6.39. Generally, a steep power-law index (i.e., high values of $n$) imply low warm gas mass fractions; we highlight, however, that the work presented here with the application of the \citet{tog16} model probes the smallest spatial scales ($<$100 pc) to this day (see \citealt{arm22} for a comparable study). Overall, our analysis shows relatively high T$_l$ for all four regions, with the highest value observed in region 2. The high T$_l$ in region 2, supports a scenario where the molecular gas has been recently heated by shocks (see Section \ref{sec:oiv}). \par
Using the model-derived parameters listed in Table \ref{table:h2_model} we estimate warm (T$_l \gtrsim$ 250 K) molecular gas masses of 3.34 $\times$ 10$^{4}$ M$_{\odot}$, 3.22 $\times$ 10$^{4}$ M$_{\odot}$, 8.76 $\times$ 10$^{3}$ M$_{\odot}$ and 1.32 $\times$ 10$^{4}$ M$_{\odot}$, for regions 1, 2, 3, and 4, respectively. \citet{tog16} fitted continuous temperature distributions to a sample of SFGs with reliable molecular gas masses and calibrated an extrapolating temperature (50 K) providing a model that can be used to infer the total molecular gas mass directly from the H$_2$ rotational emission. Similar to the approach by \citet{tog16}, using the model-derived values we extrapolate the power law distribution to a temperature of 50 K to obtain an approximate value of the total molecular gas content accounting for the warm H$_2$ component. We list the estimated total molecular masses, M($>$50K)$_{\rm H_2}$, in the fifth column in Table \ref{table:h2_model}. Our inferred H$_2$ masses indicate that a large fraction of the total H$_2$ gas mass is found in the north-west region (region 2) M($>$50 K)$_{\rm H_2}$ = 40.3 $\times$ 10$^{6}$ M$_{\odot}$.\par
The H$_2$ masses listed in the the fifth column in Table \ref{table:h2_model} account for the molecular gas across a broad range of temperatures, including the warm component. Using the $^{12}$CO (1--0) intensity map shown in Figure \ref{fig:CO_H2} we integrate the line profiles for the same four MIRI/MRS regions studied here. For the nuclear regions in M83, \citet{hir18} reports $^{12}$CO (1--0)  peak temperatures between 10 and 20 K. Assuming a Galactic $\alpha_{CO}$ factor \citep{bol13} we estimate total H$_2$ masses (probing the cold molecular gas) for all four regions and list them in the sixth column of Table \ref{table:h2_model} under M(CO)$_{\rm H_2}$. As noted above, studies have confirmed that the $\alpha_{CO}$ factor varies with metallicity. According to \citet{bol13}, for a system with a metallicity similar to that measured for the nucleus of M83, log(Z/Z$\odot$) $\sim$ $+$0.2 dex \citep{her21}, the $\alpha_{CO}$ factor should be slightly lower than the Galactic value (we note, however, the large spread in the measurements), resulting in even lower M(CO)$_{\rm H_2}$ values than those listed in Table \ref{table:h2_model}. \par

Comparing the total molecular gas masses inferred through the H$_2$ rotational lines, M($>$50 K)$_{\rm H_2}$, against those obtained using CO as a tracer of H$_2$, M(CO)$_{\rm H_2}$, we find that in all four regions the molecular gas masses accounting for the warm molecular component are higher than the M(CO)$_{\rm H_2}$ masses. These differences in masses can be attributed to the CO-dark component not probed by the CO emission. In the last column of Table \ref{table:h2_model} we list the difference between the two mass estimates, M($>$50 K)$_{\rm H_2}$ $-$ M(CO)$_{\rm H_2}$ $=$ M(CO-dark). The most extreme case appears to be observed in region 3, where M($>$50 K)$_{\rm H_2}$ is a factor of $\sim$12 higher than M(CO)$_{\rm H_2}$. In this region we estimate that $\sim$92\% of the total molecular gas mass is contained in the warm (CO-dark) component. We highlight that the high CO-dark gas fraction observed in region 3 appears to be primarily driven by the low contents of CO (see Figure \ref{fig:CO_H2}). For region 2  we estimate that $\sim$82\% of the total H$_2$ gas mass is not traced by the cold CO gas. Interestingly, the lowest fraction of CO-dark gas is found in region 1, which encapsulates the optical nucleus of this galaxy, with only $\sim$27\% of the total H$_2$ gas mass found in the warm component. 
Lastly, we note that the molecular gas in region 2 accounts for $\sim$60\% of the total estimated H$_2$ gas mass of 67.90 ($\pm$5.43) $\times$ 10$^{6}$ M$_{\odot}$ in the core of M83. \par
The spatially resolved study of the molecular gas content in the core of M83 presented here, began to confirm the predictions by \citet{her21} where they propose that the high column densities of \ion{S}{2} arise from a clumpy molecular gas environment with large quantities of CO-dark gas. The four pointings in \citet{her21} exhibiting this excess of \ion{S}{2} are located in regions 2 and 4 of the present study.  Figure \ref{fig:CO_H2} highlights the clumpy nature of the warm molecular gas, particularly in these two regions. Furthermore, for the two specific regions where the excess of \ion{S}{2} is observed, our study finds that 82\% and 73\% of the total H$_2$ gas mass (for regions 2 and 4, respectively) is contained in the warm molecular gas component.\par
Overall, this study highlights the importance of accounting for the warm molecular gas component when estimating total masses of the H$_2$ gas, specifically in regions of high SF. Our findings indicate that $\sim$75\% of the total molecular gas mass is contained in the warm component, when the fraction of CO-estimated H$_2$ mass, M(CO)$_{\rm H_{2}}$,  is eliminated from the power-law derived mass estimates, M($>$50 K)$_{\rm H_{2}}$. Past studies have shown that the CO-dark gas can account for $>$70\% of the total molecular gas, with the highest percentages found in low metallicity environments \citep{bal17, mad20,leb22,viz22}. Most recently, similar results were reported for the nearby dwarf galaxy WLM (with a metallicity of 13\% solar) where \citet{arch22} find that 90\% to 100\% of the total molecular gas mass is contained in CO-dark H$_2$ clouds. Although outside of the scope of this work, future analysis of the MIRI/MRS observations will include a comparison of the H$_2$ emission maps with the $^{12}$CO (1--0) intensity maps by \citet{hir18} to discern the spatial differences between the warm and cold molecular gas. \par

\begin{table*}
\caption{Measured H$_2$ fluxes}
\label{table:h2}
\centering 
\begin{tabular}{lrrrr}
\hline \hline
Line & Region 1 & Region 2 & Region 3 & Region 4 \\
\hline \hline
\multicolumn{5}{c}{Flux}\\
\multicolumn{5}{c}{(10$^{-18}$ $W\; m^{-2}$)}\\
\hline \hline
H$_2$ S(7) & 17.30$\pm$0.93 & 9.42$\pm$1.13& 1.66$\pm$0.76 & 5.01$\pm$0.47\\
H$_2$ S(6) & 8.07$\pm$0.20 & 5.57$\pm$0.15 & 0.92$\pm$0.12 & 2.38$\pm$0.14\\
H$_2$ S(5) & 36.99$\pm$0.92 & 31.32$\pm$2.07 & 3.91$\pm$0.64 & 9.95$\pm$0.38\\
H$_2$ S(4) & 18.32$\pm$3.06& 19.13$\pm$4.34 & 2.50$\pm$0.17& 5.45$\pm$3.36\\
H$_2$ S(3) &74.27$\pm$0.26 & 31.24$\pm$0.27& 7.05$\pm$0.11& 2.11$\pm$0.16\\
H$_2$ S(2) & 29.66$\pm$0.41 & 38.06$\pm$0.02& 7.36$\pm$0.24& 12.47$\pm$0.42\\
H$_2$ S(1) & 49.60$\pm$0.58 & 60.38$\pm$1.62 & 13.35$\pm$0.14& 20.72$\pm$0.35\\
\hline \hline
\end{tabular}
\end{table*}

\begin{table*}
\caption{H$_2$ Model-derived parameters for the center of M83}
\label{table:h2_model}
\centering 
\begin{tabular}{ccccccc}
\hline \hline
Region  & $n$ & T$_l$  & M($>$T$_l$)$_{\rm H_2}$  & M($>$50 K)$_{\rm H_2}$  & M(CO)$_{\rm H_2}$& M(CO-dark)$_{\rm H_2}$ \\
 &  & (K) & (M$_{\odot}$)  & (M$_{\odot}$) & (M$_{\odot}$) & (M$_{\odot}$)\\
\hline \hline
Region 1& 5.17$\pm$0.04  & 252$\pm$3 & 3.34 ($\pm$0.13) $\times$ 10$^{4}$ & 9.88 ($\pm$0.46) $\times$ 10$^{6}$ & 7.22 $\times$ 10$^{6}$ &  2.65 $\times$ 10$^{6}$\\
Region 2 & 6.22$\pm$0.11 & 301$\pm$7 &  3.22 ($\pm$0.25) $\times$ 10$^{4}$ & 40.3 ($\pm$4.92) $\times$ 10$^{6}$ &  7.06 $\times$ 10$^{6}$ &  33.2 $\times$ 10$^{6}$\\
Region 3 & 6.33$\pm$0.18 & 269$\pm$9 &  8.76 ($\pm$0.82) $\times$ 10$^{3}$ & 10.0 ($\pm$2.00) $\times$ 10$^{6}$ &  0.80 $\times$ 10$^{6}$ & 9.22 $\times$ 10$^{6}$\\
Region 4 & 5.70$\pm$0.11 & 267$\pm$9 & 1.32 ($\pm$0.14) $\times$ 10$^{4}$ & 7.72 ($\pm$1.03) $\times$ 10$^{6}$ &  2.07 $\times$ 10$^{6}$ &  5.65 $\times$ 10$^{6}$\\
\hline
Total & & & 8.76 ($\pm0.33$) $\times$ 10$^{4}$ &67.90 ($\pm 5.43$) $\times$ 10$^{6}$ & 17.15 $\times$ 10$^{6}$ &  50.74 $\times$ 10$^{6}$\\
\hline \hline

\end{tabular}
\end{table*}

             \begin{figure}
   	  \centerline{\includegraphics[scale=0.75]{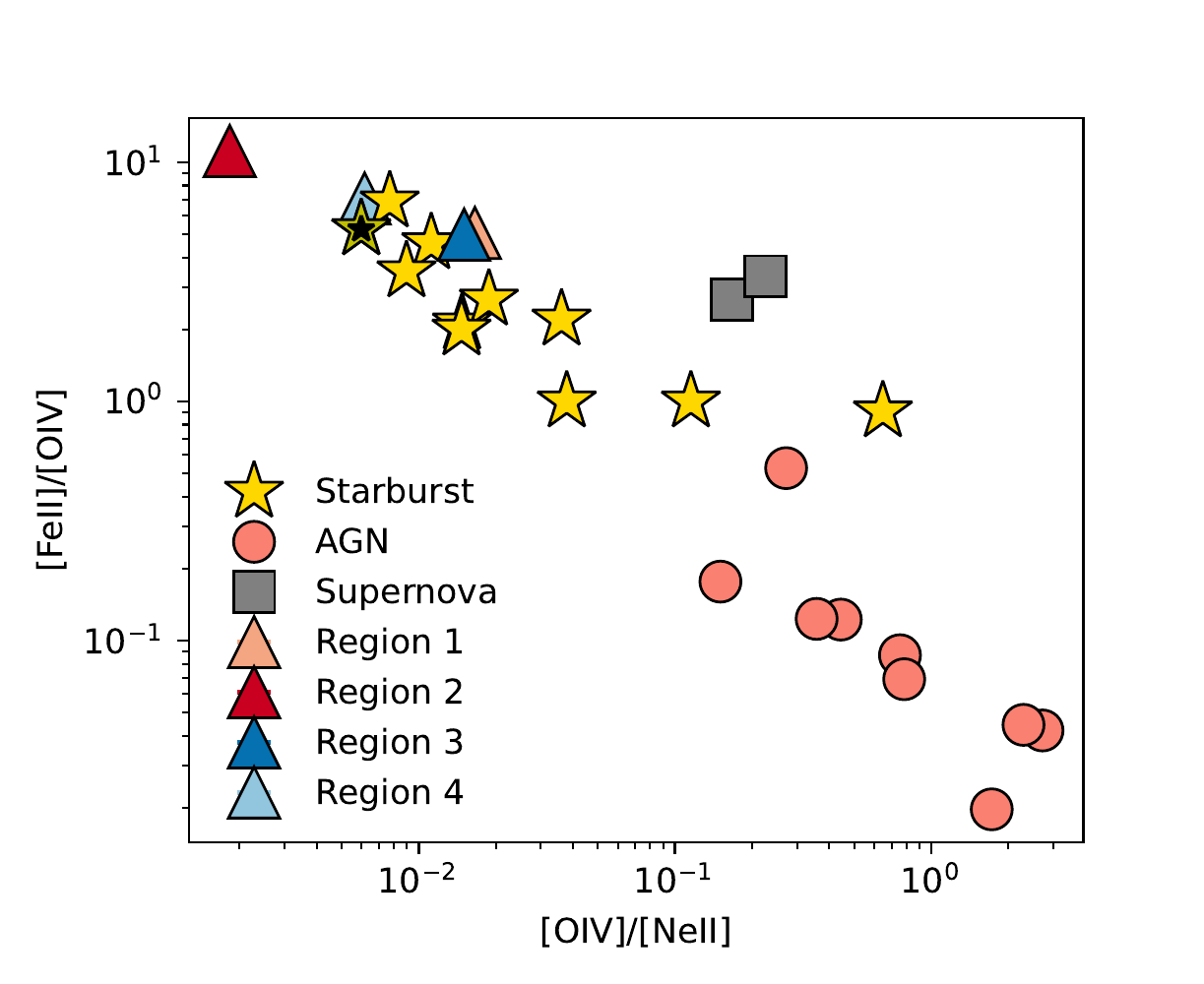}}
      \caption{Fine-structure emission line ratios for the four different regions in the nucleus of M83 (shown in triangles). We compare these ratios with typical values obtained from sources with starburst-dominated (yellow stars from \citealt{ver03}), AGN-dominated (salmon circles from \citealt{stu02}), and SNR (grey squares from \citealt{oli99, oli99b}) emission.}
         \label{fig:shocks}
   \end{figure}

             \begin{figure}
   	  \centerline{\includegraphics[scale=0.48]{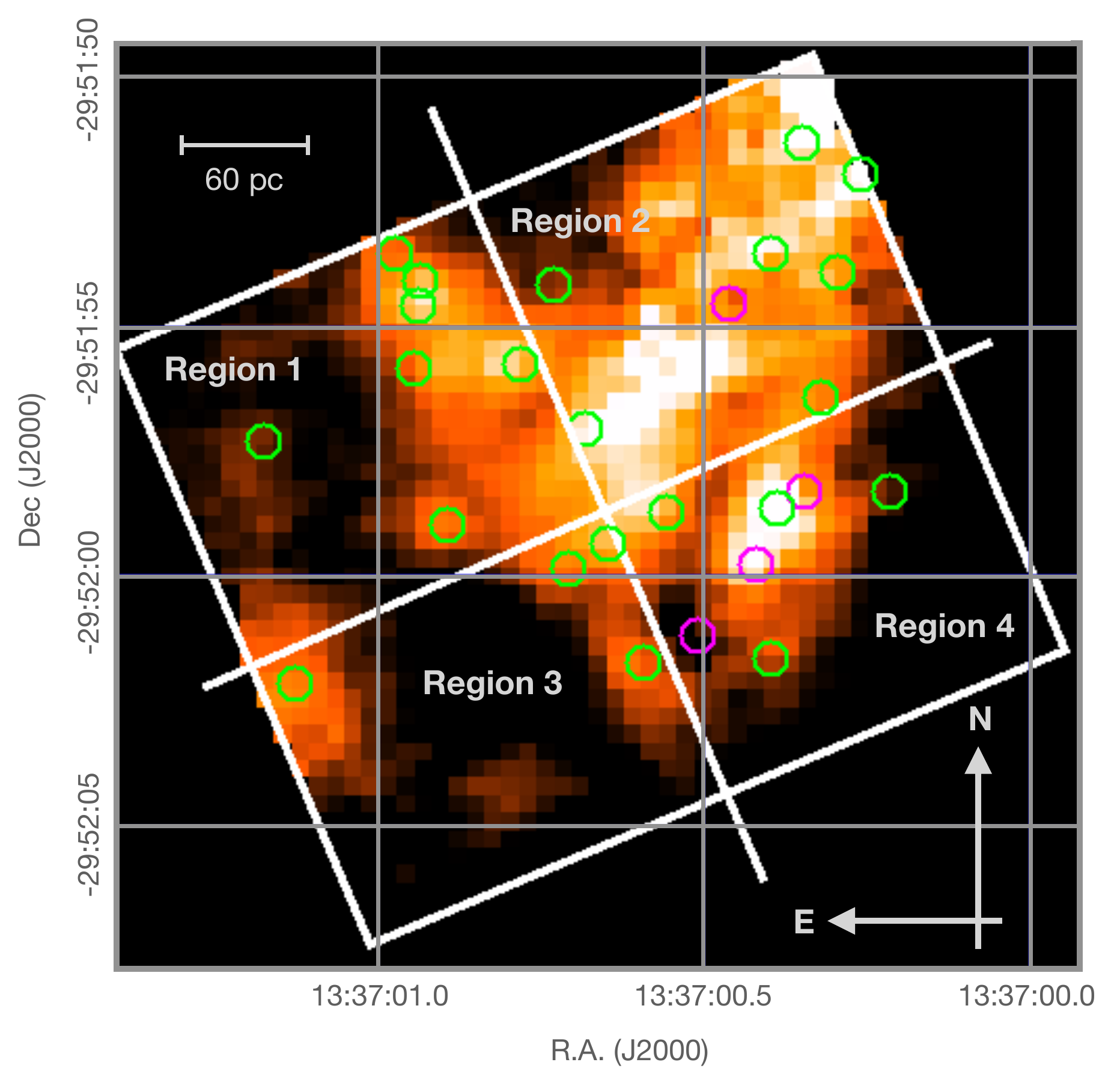}}
      \caption{[FeII] emission line map as observed by MIRI/MRS channel 4. We label the four different regions accordingly. We show with green apertures the location of supernovae remnants by \citet{win22}. We display with magenta apertures the location of four young massive clusters studied in \citet{her21}.}
         \label{fig:feII}
   \end{figure}

\subsection{[\ion{O}{4}] 25.89$\mu$m emission and possible presence of shocks}\label{sec:oiv}
Given that hot, massive stars emit a limited number of ionizing photons beyond the He$^+$ edge at 54 eV, the spectra of starburst sources are typically dominated by low-excitation emission lines. The excitation of high-ionization species, on the other hand, requires extreme conditions. These high-excitation species include the [\ion{O}{4}] 25.89$\mu$m line with an ionization potential just above the He$^+$ edge. Surprisingly, past studies have detected this emission line in several different starbursts \citep[e.g., ][]{lut98, ver03, hao09, ber09}. For many of these objects, the [\ion{O}{4}] detection cannot be attributed to an enshrouded AGN; instead the mechanism believed to be responsible for the production of these highly ionizing photons ($\gtrsim$54 eV) is very hot stars (e.g., Wolf-Rayet stars), ionizing shocks \citep{lut98}, and although debated, most recently it was suggested that high-mass X-ray binaries could also be responsible for these photons \citep{sch19, sen20}.  We note that no AGN has been identified in M83, and even the possible presence of a massive black hole is debated \citep[cf.][Appendix A2]{rus20}. However, some hard X-ray point sources are seen (at least in projection) toward the nuclear region \citep{lon14, rus20}.\par

In Figure \ref{fig:m83_oiv_feii} we show the emission profiles of [\ion{O}{4}] 25.89$\mu$m and [\ion{Fe}{2}] 25.99$\mu$m. Due to the strong H$_2$ emission discussed in the previous section, it is conceivable that ionizing shocks are indeed contributing to the observed [\ion{O}{4}] emission. One way to explore the possible shock contributions in these active environments is to compare the strength of the [\ion{O}{4}] feature to the emission from other fine-structure lines expected to be strong in partially ionized regions with shocks.\par
Since Fe is known to strongly deplete onto grains in the ISM \citep{sav96}, the shock destruction of dust grains typically boosts the [\ion{Fe}{2}] emission (already present in ionized zones). The [\ion{Fe}{2}]/[\ion{O}{4}] ratio along with the [\ion{O}{4}]/[\ion{Ne}{2}] fluxes are typically used to distinguish between excitations driven by starbursts, AGN or supernova remnant shocks \citep[e.g., ][]{lut03, stu06, ina13}. In Figure \ref{fig:shocks} we show the fine-structure emission line ratios from starburst-dominated systems measured by \citet{ver03} as yellow stars, along with the ratios inferred for AGN-dominated emission by \citet{stu02} as salmon-colored circles, and line ratios from supernova remnants in grey squares \citep{oli99, oli99b}. Both starburst-dominated systems and supernova remnant shocks tend to show much weaker [\ion{O}{4}] emission (e.g., higher [\ion{Fe}{2}]/[\ion{O}{4}] ratios), compared to those from AGNs, therefore occupying the top half of Figure \ref{fig:shocks}. \par
The [\ion{O}{4}] emission from the nuclear starburst in M83 was previously detected in observations taken with ISO Short Wavelength Spectrometer (ISO-SWS) and reported by \citet{ver03}, marked with a black star in Figure \ref{fig:shocks}. We also include in Figure \ref{fig:shocks} the inferred ratios for the four pre-defined regions in the core of M83, shown as triangles. We highlight that the fine-structure emission line ratios reported by \citet{ver03} for M83 are in essence the integrated fluxes for all of the four regions in this study. From Figure \ref{fig:shocks} it is clear that all four regions in the core of M83 have comparable ratios to those characteristic of starburst systems with variations from region to region, and similarly weak [\ion{O}{4}] emission to those detected in supernova remnant shocks. \par

Due to the nature of iron, as detailed earlier, previous studies have extensively used the near-infrared [\ion{Fe}{2}] emission in starburst galaxies to identify new supernova remnants \citep[SNRs; e.g.,][]{lab06}. Furthermore, according to \cite{bla14} and \citet{win22}, a number of the known SNRs in M83 have indeed been identified as strong [\ion{Fe}{2}] emitters. We show in Figure \ref{fig:feII} the [\ion{Fe}{2}] emission at 25.99 $\micron$. To investigate if the [\ion{Fe}{2}] emission displays any structure coincident with SNRs, we highlight in Figure \ref{fig:feII} with green apertures the location of SNRs most recently identified in the nuclear region by \citet{win22}, many of which are isolated [\ion{Fe}{2}] sources. The [\ion{Fe}{2}] emission appears to be diffuse and mostly concentrated in region 2, extending into regions 1 and 4. Most of the SNRs are coincident with the region of bright diffuse [\ion{Fe}{2}] emission, but enhanced emission from the individual SNRs in not generally apparent.  \par
Studies such as that by \citet{lab06} have reported [\ion{Fe}{2}] emission with a spatially-extended component in several nearby starburst galaxies, where the SNR [\ion{Fe}{2}] emission does indeed account for $\sim$10\% of the total [\ion{Fe}{2}] luminosity. We also show in Figure \ref{fig:feII} with magenta apertures the location of four young massive clusters studied in \citet{her21} with ages ranging from $\sim$ 3 to 10 Myr \citep{wof11,her19}. Two of these clusters appear to bracket a region with observed strong [\ion{Fe}{2}] emission while two others do not have a strong association, making an attribution unclear. \par

\begin{figure*}
   	  \centerline{\includegraphics[scale=0.32]{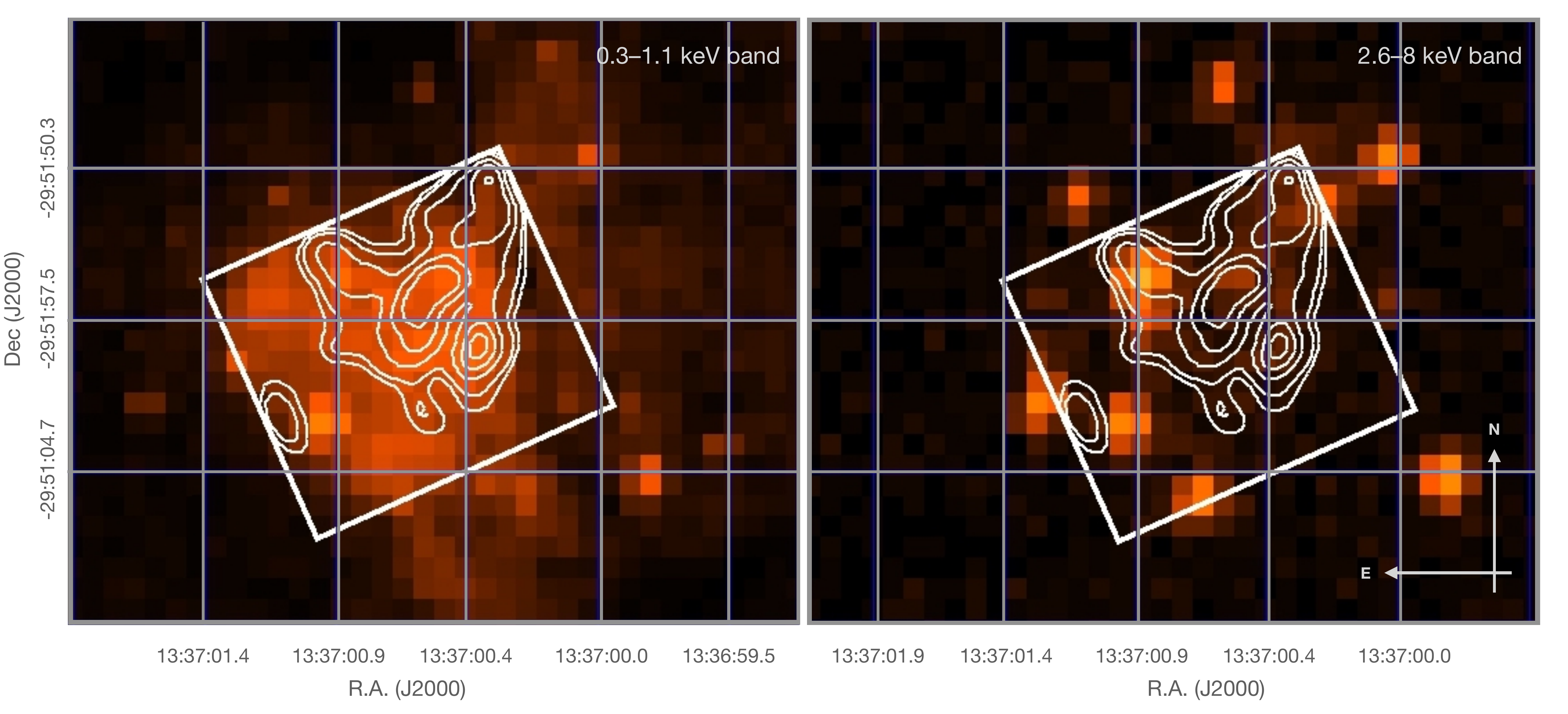}}
      \caption{In both panels we show the [\ion{Fe}{2}] 25.99 $\micron$ emission contours in white, along with the MIRI/MRS channel 4 FoV for context. \textit{Left: } Chandra image in the 0.3-1.1 keV band by \citet{lon14}. \textit{Right: } Chandra image in the 2.6-8 keV band by \citet{lon14}. }
         \label{fig:xray}
   \end{figure*}

As mentioned before, the nucleus of M83 has been studied extensively at different wavelengths. For example, recently \citet{rus20} presented a detailed discussion of the multiwavelength characteristics of the nucleus (their Section 4.5 and Figure 11). Similarly, \citet{lon14} have carefully studied the X-ray emission in the nucleus of M83. These authors highlight that the soft X-ray emission (0.35-1.1 keV) in the nuclear region of the galaxy has a clumpy and diffuse nature, while the hard X-ray emission (2.6-8.0 keV) is dominated by point sources which are primarily high mass X-ray binaries. The soft X-ray emission in the nuclear region, they propose, is energized by both the SNR population and the strong stellar winds from all the young massive stars.  In Figure \ref{fig:xray} we compare the \ion{Fe}{2} emission contours detected in the MIRI observations, to the soft and hard X-ray emission maps by \citet{lon14}. Overall, the observed diffuse soft X-ray emission is comparable in nature to the [\ion{Fe}{2}] emission, this in contrast to the point-source emission from hard X-rays. Interestingly, \citet{zha18b} reviews the nature of galactic winds and stellar feedback from starbursts and describes a scenario where the starburst region launches a hot wind that expands outward. As this hot wind encounters the
surrounding large-scale ISM, shock waves produce structured soft X-ray emission. Under this scenario the observed clumpy and diffuse soft X-ray emission, as well as the similarly diffuse [\ion{Fe}{2}] emission, could be tracing the shocks from such an interaction. \par 

             \begin{figure}
   	  \centerline{\includegraphics[scale=0.3]{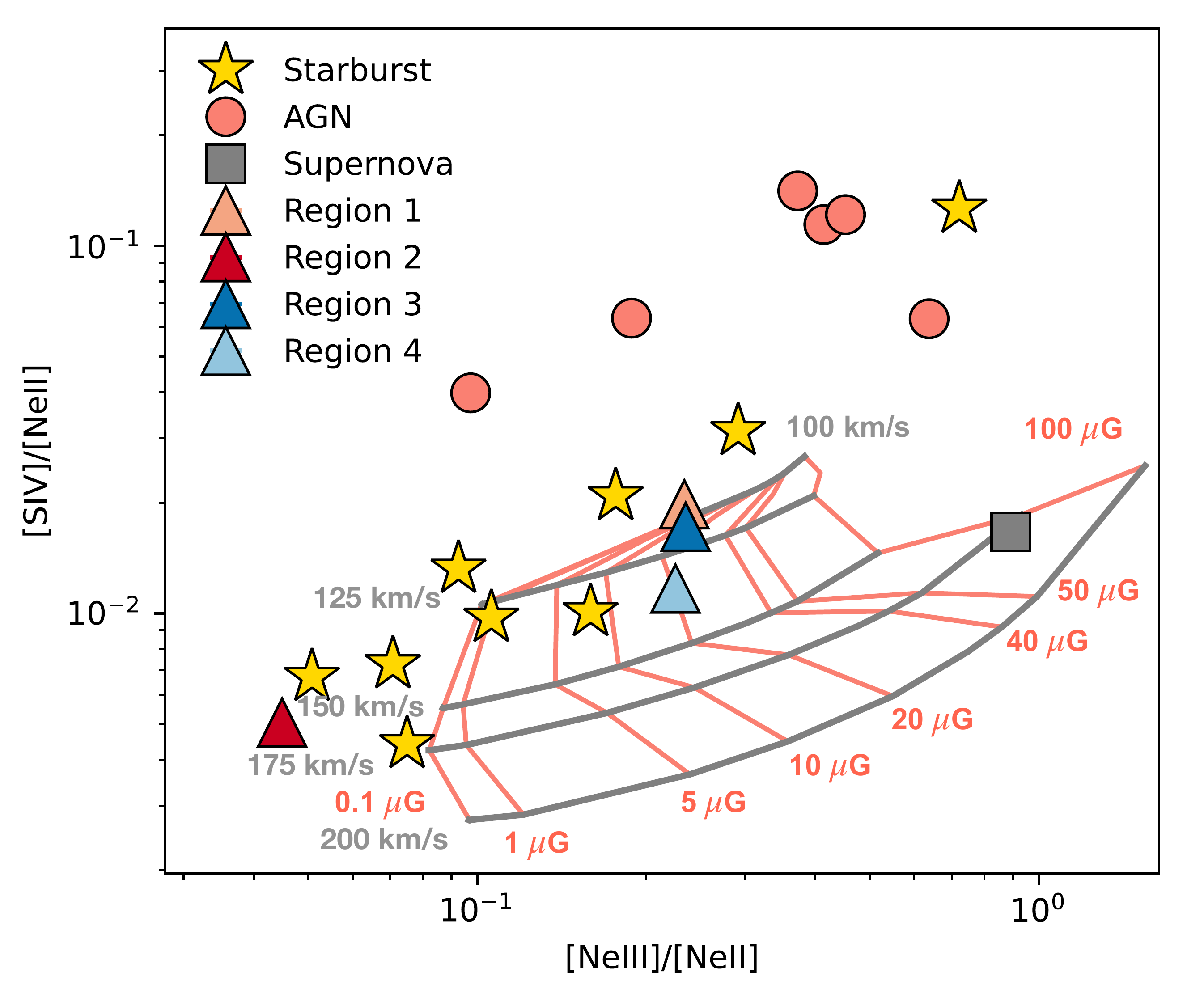}}
      \caption{ [SIV]/[NeII] vs. [NeIII]/[NeII] emission line diagnostic diagram. The ratios for the different regions are shown with triangles of different colors. We show in grey the shock speeds and in salmon the magnetic field strengths according to the shock ionization model grid by \citet{all08}. For completion, we also include the typical values from sources with starburst-dominated (yellow stars from \citealt{ver03}), AGN-dominated (salmon circles from \citealt{stu02}), and SNR (grey squares from \citealt{oli99, oli99b}) emission.}
         \label{fig:shocks_models}
   \end{figure}

Both photoionization and shocks can in principle excite gas in starburst systems. Assuming a scenario  similar to that described in \citet{zha18b}, we can explore the properties of the expected shocks. In Figure \ref{fig:shocks_models} we show the [\ion{S}{4}]/[\ion{Ne}{2}] versus [\ion{Ne}{3}]/[\ion{Ne}{2}] emission line diagnostic diagram and overlay the shock ionization model grids of \citet{all08}, with magnetic field strengths ranging from 0.1 to 100 $\mu$G and shock speeds of 100 to 200 km s$^{-1}$. According to these shock ionization models and assuming shocks are the main ionization mechanism in these intense environments, the shocks observed in regions 1, 3 and 4 display speeds of $\lesssim$ 150 km s$^{-1}$, whereas the shock speeds observed in region 2 is $>$ 150 km s$^{-1}$. Additionally, these models indicate that the strength of the magnetic fields is similar in regions 1, 3 and 4 ($\sim$ 20 $\mu$G), and higher in these regions than in region 2 ($\ll$0.1 $\mu$G). For completion, we also include in Figure \ref{fig:shocks_models} the line ratios of starburst-, AGN- and supernova-dominated emission. We note that for a fraction of the objects in the starburst-dominated sample, the observed ratios agree with the shock ionization model by \citet{all08}. We stress that while the emission line ratios measured in the different nuclear regions in M83 are consistent with values expected from shock ionization, comparing to photoionization models by \citet{lev10} we confirmed that young starbursts with ages $\sim$3 Myr, above-solar metallicity and ionization parameter of 2 $\times$ 10$^7$ cm s$^{-1}$ can similarly reproduce the measured flux ratios. A similar trend is observed for the fraction of objects in the starburst-dominated sample that agree with the shock ionization models; these ratios can also be reproduced by a young starburst. Definitive confirmation of a shock-dominated environment in the nucleus of M83 will require self-consistent modeling of the ionization by shocks and massive stars, a task that will be explored in future studies. \par

\section{Summary} \label{sec:summary}
We present the first analysis of the recently-acquired JWST MIRI/MRS observations of the nucleus of M83, covering a region of $\sim$200 pc $\times$ 200 pc. The spectroscopic observations exhibit a plethora of emission features characterizing the multi-phase ISM and its different components: ionized gas, warm molecular gas, and dust. We summarize our findings as follows:
\begin{itemize}
\item Assuming a uniform power law model as proposed by \citet{tog16}, we estimate for the first time molecular gas masses accounting for the warm H$_2$ gas component in four regions in the core of M83 obtaining values of 9.88 $\times$ 10$^{6}$ M$_{\odot}$, 40.3 $\times$ 10$^{6}$ M$_{\odot}$, 10.0 $\times$ 10$^{6}$ M$_{\odot}$ and 7.72 $\times$ 10$^{6}$ M$_{\odot}$, for regions 1, 2, 3, and 4, respectively.
\item We report on the molecular gas masses as traced by the $^{12}$CO (1--0) transition spatially coincident with the four MIRI/MRS regions in the nucleus of M83. We find that in regions 2, 3 and 4, the CO-dark gas accounts for $>$70 \% of their total molecular gas masses.
\item We find that the total molecular gas mass accounting for the warm H$_2$ component inferred in region 3 (south of the optical nucleus) is a factor of $\sim$12 higher than the mass inferred through the CO emission. This difference in molecular gas masses appears to be driven by a localized lack of cold molecular gas emission as traced by the $^{12}$CO (1--0) transition. 
\item Our findings indicate that $\sim$75\% of the total molecular gas mass is contained in the warm H$_2$ component.
\item We detect emission from the high-excitation species [OIV] 25.89$\micron$ in the core of M83 and report varying emission strengths in all four regions.
\item We compare the [FeII] 25.99 $\mu$m emission map to the location of SNRs \citep{win22} and a few massive stellar clusters \citep{her21} and find no correlation between the [\ion{Fe}{2}] emission and the location of these sources. We instead observe clumpy and diffuse [FeII] 25.99 $\mu$m emission, comparable to the soft X-Ray emission detected in the nucleus of this galaxy by \citet{lon14}.
\item We propose that the diffused [\ion{Fe}{2}] 25.99 $\mu$m emission might be tracing shocks created during the interactions between the hot wind produced by the starburst and the much cooler ISM above the galactic plane, similar to the model described in \citet{zha18b}. Confirmation of this scenario requires self-consistent photoionization and shock modeling.  

\end{itemize}
The initial study presented here shows the incredible power and bright future of this new mid-infrared instrument and configuration, MIRI/MRS. We highlight that we will further exploit these MRS observations focusing on spatially-resolved studies of the physical and chemical properties of the multi-phase ISM in the nuclear region of M83 (Jones et al. in prep.). The spectroscopic capabilities and unprecedented mid-infrared sensitivity of MIRI are allowing us to better understand the intense environments of these nearby starburst galaxies, in many cases comparable to those expected in the very first galaxies.

\begin{acknowledgments}
We are grateful to Akihiko Hirota for sharing their calibrated  Atacama Large Millimeter/submillimeter Array (ALMA) CO emission map. This work is based on observations made with the NASA/ESA/CSA \textit{James Webb Space Telescope}. Support for program JWST-GO-02291 was provided by NASA through a grant from the Space Telescope Science Institute, which is operated by the Associations of Universities for Research in Astronomy, Incorporated, under NASA contract NAS5-26555. This work is based on observations made with the NASA/ESA/CSA James Webb Space Telescope. The data were obtained from the Mikulski Archive for Space Telescopes at the Space Telescope Science Institute, which is operated by the Association of Universities for Research in Astronomy, Inc., under NASA contract NAS 5-03127 for JWST. These observations are associated with program \#02219.
\end{acknowledgments}

%

\vspace{5mm}
\facilities{JWST(MIRI)}


\software{dust\_extinction \citep{gor21}
          }



\appendix

\section{Appendix information}\label{sec:app_excitation}
\begin{figure}
   	  \centerline{\includegraphics[scale=0.45]{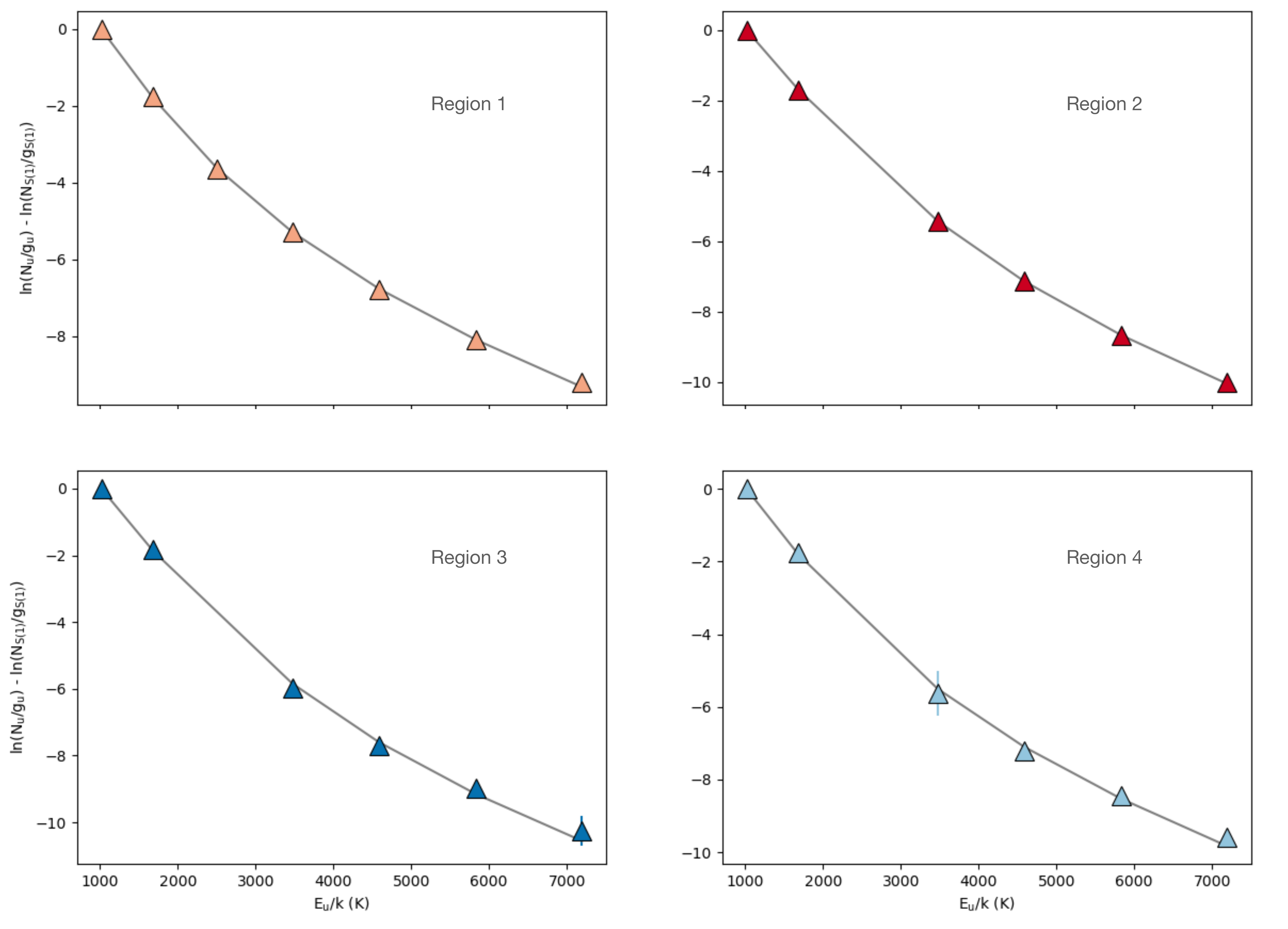}}
      \caption{Excitation diagrams for all four M83 nuclear regions. Similar to the work by \citet{tog16} the N$_u$/g$_u$ ratios are normalized to the S(1) transition. We show with a grey line the best model fit adopting the parameters listed in Table \ref{table:h2_model}.} 
         \label{fig:excite}
   \end{figure}

The H$_2$ gas masses were inferred adopting the approach by \citet{tog16}. We modeled the H$_2$ excitation and obtained warm H$_2$ gas masses (M($>$T$_l$)$_{\rm H_2}$; fourth column in Table \ref{table:h2_model}), by varying the power-law slope and T$_l$. In Figure \ref{fig:excite} we show the excitation diagrams for the different regions in the nucleus of M83. As noted by \citet{don23}, the S(3) transition at 9.665 $\micron$ is found at the peak of the silicate absorption, and therefore highly sensitive to extinction. In the analysis by \citet{don23}, they include this line with high uncertainties to ensure it does not bias their H$_2$ excitation model fits. For our analysis, we exclude the S(3) transitions for Regions 2, 3 and 4, as they appear to strongly bias the model fits, indicating high degrees of extinction. Overall, the best model fits accurately characterize the observed columns densities.\par
As detailed in Section \ref{sec:h2}, the inferred models recover the total molecular gas mass, M($>$50 K)$_{\rm H_2}$, by extrapolating the fitted power-law temperature distributions to the calibrated single lower cutoff temperature \citep[50 K;][]{tog16}.

\bibliography{M83_MIRI}{}
\bibliographystyle{aasjournal}



\end{document}